\documentclass{article}
\pdfoutput=1

\usepackage{amsmath,amssymb,feynmp-auto}
\usepackage{fullpage}

\DeclareMathOperator{\ud}{d\!}

\begin{document}

\title{No-go for tree-level R-symmetry breaking}
\author{Feihu Liu\textsuperscript{a, *},
        Muyang Liu\textsuperscript{b, \dag} and
        Zheng Sun\textsuperscript{b, c, \ddag}\\
        \textsuperscript{a}%
        \normalsize\textit{School of Physical Electronics, University of Electronic Science and Technology of China,}\\
        \normalsize\textit{Chengdu 610054, P. R. China}\\
        \textsuperscript{b}%
        \normalsize\textit{Center for Theoretical Physics, College of Physical Science and Technology,}\\
        \normalsize\textit{Sichuan University, Chengdu 610064, P. R. China}\\
        \textsuperscript{c}%
        \normalsize\textit{CAS Key Laboratory of Theoretical Physics, Institute of Theoretical Physics,}\\
        \normalsize\textit{Chinese Academy of Sciences, Beijing 100190, P. R. China}\\
        \normalsize\textit{E-mail:}
        \textsuperscript{*}\texttt{liufeihu04@gmail.com,}
        \textsuperscript{\dag}\texttt{liumuyang14@hotmail.com,}
        \textsuperscript{\ddag}\texttt{sun\_ctp@scu.edu.cn}
        }
\date{}
\maketitle

\begin{abstract}

We show that in gauge mediation models with tree-level R-symmetry breaking where supersymmetry and R-symmetries are broken by different fields, the gaugino mass either vanishes at one loop or finds a contribution from loop-level R-symmetry breaking.  Thus tree-level R-symmetry breaking for phenomenology is either no-go or redundant in the simplest type of models.  Including explicit messenger mass terms in the superpotential with a particular R-charge arrangement is helpful to bypass the no-go theorem, and the resulting gaugino mass is suppressed by the messenger mass scale.

\end{abstract}

\section{Introduction}

Supersymmetry (SUSY)~\cite{Martin:1997ns, Quevedo:2010ui} provides a natural solution to several unsolved problems in the Standard Model (SM), such as the gauge hierarchy problem, gauge coupling unification and dark matter candidates.  Since supersymmetric particles (sparticles) have not been discovered yet, SUSY must be broken to give them heavy masses escaping the current experimental limit.  To avoid light sparticles in the supersymmetric standard model (SSM), SUSY must be broken in a hidden sector, and then the SUSY breaking effects are mediated to the observable SSM sector by a messenger sector, giving sparticle mass spectrum and coupling constants which may be examined at the LHC or other future experiments.  There are three competitive mediation mechanisms:  gravity mediation, gauge mediation, and anomaly mediation.  We are focusing on gauge mediation models~\cite{Giudice:1998bp, Meade:2008wd, Kitano:2010fa} in this work.

Following the discussion of the Nelson-Seiberg theorem~\cite{Nelson:1993nf, Intriligator:2007py, Kang:2012fn} which summarizes earlier observations in Wess-Zumino and dynamical SUSY breaking models~\cite{Fayet:1975ki, Affleck:1983vc, Affleck:1984mf, Affleck:1984xz}, R-symmetries are required to build a generic SUSY breaking model.  From phenomenology point of view, the R-symmetry needs to be broken spontaneously in order to allow for the Majorana gaugino mass.  The R-symmetry is usually broken by the SUSY breaking spurion field, or pseudomodulus~\cite{Ray:2006wk, Sun:2008nh, Shih:2007av, Curtin:2012yu} which gets a vacuum expectation value (VEV) at loop level through the Coleman-Weinberg potential~\cite{Coleman:1973jx}, or through the inclusion of D-terms~\cite{Azeyanagi:2012pc, Vaknin:2014fxa}.  There are also models with tree-level R-symmetry breaking from tree-level VEVs of fields other than the pseudomodulus~\cite{Carpenter:2008wi, Sun:2008va}.  These models usually involve many fields with specific R-charges, and the gaugino mass is often generated from multiple VEVs of fields at both loop level and tree level in such complicated models~\cite{Kang:2010ye}.  A wide class of tree-level SUSY and R-symmetry breaking models with classically stable pseudomoduli spaces have been shown to give zero gaugino masses at one-loop level~\cite{Komargodski:2009jf, Abel:2009ze}.  Nevertheness, it still remains unclear whether in principle the gaugino mass could be generated just from tree-level R-symmetry breaking.

In gauge mediation models, SUSY breaking fields are coupled to messengers which are charged under the SM gauge symmetry.  SUSY breaking is mediated to the SSM sector through gauge interactions, and soft terms such as the gaugino mass emerge at low energy.  For loop-level R-symmetry breaking, the SUSY breaking spurion field $X$ also breaks the R-symmetry at loop level.  It obtains the VEV
\begin{equation} \label{eq:1-1}
    X = \langle X \rangle + \theta^2 F_X .
\end{equation}
The resulting gaugino mass at one-loop level is
\begin{equation} \label{eq:1-2}
    M_{\tilde g} \sim \frac{\alpha}{4 \pi} \frac{F_X}{\langle X \rangle} .
\end{equation}
A tree-level R-symmetry breaking model has at least two spurions which break SUSY and R-Symmetry respectively.  They have VEVs
\begin{equation} \label{eq:1-3}
    X = \theta^2 F_X , \quad
    Y = \langle Y \rangle .
\end{equation}
As we are to show, in our simplest type of models, there is no valid one-loop diagram for the gaugino mass with R-symmetries respected at all vertices, unless $X$ and $Y$ fields have identical R-charges which make the condition \eqref{eq:1-3} non-generic.  Thus tree-level R-symmetry breaking fails its original motivation to generate the gaugino mass, and we obtain a no-go statement for these models.  We are also to show that it is possible to bypass the no-go theorem by including explicit messenger mass terms in the superpotential with a particular R-charge arrangement, and the resulting gaugino mass is suppressed by the ratio between the R-symmetry breaking scale and the messenger mass scale.

\section{Gaugino masses in ordinary gauge mediation models}

\begin{figure}
        \centering 
        \begin{fmfgraph*}(100, 35)
            \fmfleft{i1}
            \fmfright{o1}
            \fmf{fermion,label=$\tilde \psi$,label.side=left}{v1,i1}
            \fmf{fermion,label=$\psi$,label.side=right}{v1,o1}
            \fmfv{decoration.shape=cross,decoration.size=5thick,label=$i \lambda \langle X \rangle$,label.angle=90}{v1}
        \end{fmfgraph*}
        \qquad
        \begin{fmfgraph*}(100, 35)
            \fmfleft{i1}
            \fmfright{o1}
            \fmf{fermion,label=$\tilde \psi^\dagger$,label.side=right}{i1,v1}
            \fmf{fermion,label=$\psi^\dagger$,label.side=left}{o1,v1}
            \fmfv{decoration.shape=cross,decoration.size=5thick,label=$i \lambda^* \langle X \rangle^*$,label.angle=90}{v1}
        \end{fmfgraph*}
        \\
        \begin{fmfgraph*}(100, 35)
            \fmfleft{i1}
            \fmfright{o1}
            \fmf{scalar,label=$\phi$,label.side=left}{v1,i1}
            \fmf{scalar,label=$\tilde \phi$,label.side=right}{v1,o1}
            \fmfv{decoration.shape=cross,decoration.size=5thick,label=$i \lambda F_X$,label.angle=90}{v1}
        \end{fmfgraph*}
        \qquad
        \begin{fmfgraph*}(100, 35)
            \fmfleft{i1}
            \fmfright{o1}
            \fmf{scalar,label=$\phi^*$,label.side=right}{i1,v1}
            \fmf{scalar,label=$\tilde \phi^*$,label.side=left}{o1,v1}
            \fmfv{decoration.shape=cross,decoration.size=5thick,label=$i \lambda^* F_X^*$,label.angle=90}{v1}
        \end{fmfgraph*}
        \\[1ex]
        \begin{fmfgraph*}(100, 70)
            \fmfleft{i1}
            \fmfright{o1,o2}
            \fmf{boson,label=$\lambda^a (\tilde g)$,label.side=right}{i1,v1}
            \fmf{fermion}{i1,v1}
            \fmf{scalar,label=$\phi^* (\tilde \phi^*)$,label.side=right}{v1,o1}
            \fmf{fermion,label=$\psi (\tilde \psi)$,label.side=right}{o2,v1}
            \fmfv{decoration.shape=circle,decoration.size=2thick,label=$- i \sqrt{2} g_a T^a$,label.angle=0,label.dist=10thin}{v1}
        \end{fmfgraph*}
        \qquad
        \begin{fmfgraph*}(100, 70)
            \fmfleft{i1}
            \fmfright{o1,o2}
            \fmf{boson,label=$\lambda^{a \dagger} (\tilde g^\dagger)$,label.side=left}{v1,i1}
            \fmf{fermion}{v1,i1}
            \fmf{scalar,label=$\phi (\tilde \phi)$,label.side=left}{o1,v1}
            \fmf{fermion,label=$\psi^\dagger (\tilde \psi^\dagger)$,label.side=left}{v1,o2}
            \fmfv{decoration.shape=circle,decoration.size=2thick,label=$- i \sqrt{2} g_a T^a$,label.angle=0,label.dist=10thin}{v1}
        \end{fmfgraph*}
        \caption{Messenger coupling vertices related to the gaugino mass.}
        \label{fg:1}
    \end{figure}
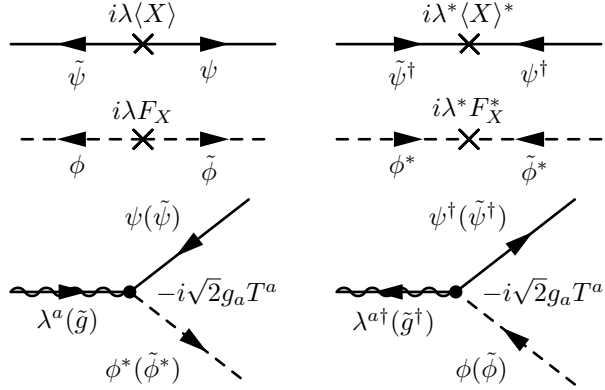

\begin{figure}
        \centering
        \begin{fmfgraph*}(120,80)
            \fmfleft{i1}
            \fmfright{o1}
            \fmf{fermion}{i1,v1}
            \fmf{photon}{i1,v1}
            \fmf{fermion}{o1,v2}
            \fmf{photon}{o1,v2}
            \fmf{phantom,left,tension=0.7,tag=1}{v1,v2}
            \fmf{phantom,right,tension=0.7,tag=2}{v1,v2}
            \fmfdot{v1,v2}
            \fmffreeze
            \fmfipath{p[]}
            \fmfiset{p1}{vpath1(__v1,__v2)}
            \fmfiset{p2}{vpath2(__v1,__v2)}
            \fmfi{fermion,label=$r - 1$}{subpath (length(p1)/2,0) of p1}
            \fmfi{fermion,label=$- r - 1$}{subpath (length(p1)/2,length(p1)) of p1}
            \fmfi{scalar,label=$r$}{subpath (0,length(p2)/2) of p2}
            \fmfi{scalar,label=$- r$}{subpath (length(p2),length(p2)/2) of p2}
            \fmfiv{decoration.shape=cross,decoration.size=5thick,label=$\langle X \rangle$,label.angle=90}{point length(p1)/2 of p1}
            \fmfiv{decoration.shape=cross,decoration.size=5thick,label=$F_X^*$,label.angle=-90}{point length(p2)/2 of p2}
        \end{fmfgraph*}
        \qquad
        \begin{fmfgraph*}(120,80)
            \fmfleft{i1}
            \fmfright{o1}
            \fmf{fermion}{v1,i1}
            \fmf{photon}{v1,i1}
            \fmf{fermion}{v2,o1}
            \fmf{photon}{v2,o1}
            \fmf{phantom,left,tension=0.7,tag=1}{v1,v2}
            \fmf{phantom,right,tension=0.7,tag=2}{v1,v2}
            \fmfdot{v1,v2}
            \fmffreeze
            \fmfipath{p[]}
            \fmfiset{p1}{vpath1(__v1,__v2)}
            \fmfiset{p2}{vpath2(__v1,__v2)}
            \fmfi{fermion,label=$r - 1$}{subpath (0,length(p1)/2) of p1}
            \fmfi{fermion,label=$- r - 1$}{subpath (length(p1),length(p1)/2) of p1}
            \fmfi{scalar,label=$r$}{subpath (length(p2)/2,0) of p2}
            \fmfi{scalar,label=$- r$}{subpath (length(p2)/2,length(p2)) of p2}
            \fmfiv{decoration.shape=cross,decoration.size=5thick,label=$\langle X \rangle^*$,label.angle=90}{point length(p1)/2 of p1}
            \fmfiv{decoration.shape=cross,decoration.size=5thick,label=$F_X$,label.angle=-90}{point length(p2)/2 of p2}
        \end{fmfgraph*}
        \caption{One-loop diagrams for the gaugino mass in ordinary gauge mediation models, with R-charges of messenger components determined from R-charge conservation.  The two diagrams are conjugate to each other, and both should be included as Majorana mass terms.}
        \label{fg:2}
    \end{figure}
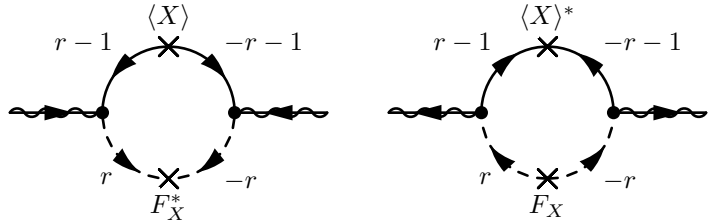

We will review some result of gauge mediation and set up the notations for our following analysis.  We start from the superfield formulation of the standard SUSY Lagrangian
\begin{equation}
    L = L_\text{Kinetic} + [W]_{\theta \theta} + \text{c.c.} ,
\end{equation}
and expand it in the component field formulation.  Since we are concerning how gauginos acquire masses after SUSY and R-symmetry breaking, we ignore the detail of the SUSY breaking sector, and assume a spurion field $X$ as specified in \eqref{eq:1-1}.  $X$ couples to the messenger sector through the cubic term in the superpotential
\begin{equation}
    W = \lambda X \tilde \Phi \Phi ,
\end{equation}
where $\tilde \Phi$ and $\Phi$ are messengers which are conjugate to each other in SM gauge symmetry representations.  The non-zero VEV of $X$ gives a SUSY breaking spectrum to messenger fields, which can be seen from expanding the SUSY lagrangian in component fields:
\begin{equation}
    [\lambda X \tilde \Phi \Phi]_{\theta \theta} + \text{c.c.} = \lambda \langle X \rangle \tilde \psi \psi + \lambda^* \langle X \rangle^* \tilde \psi^\dagger \psi^\dagger + \lambda F_X \tilde \phi \phi + \lambda^* F_X^* \tilde \phi^* \phi^* + \dotsb .
\end{equation}
Messengers are charged under SM gauge symmetry, thus coupled to gauge fields through covariant derivative terms in the kinetic part of the Lagrangian.  In the Wess-Zumino gauge, the contact terms of messengers to gauginos are
\begin{equation}
    [\Phi^\dagger (e^{2 g_a T^a V}) \Phi]_{\theta \theta \bar \theta \bar \theta} = - \sqrt{2} g_a (\phi^* T^a \psi) \lambda^a - \sqrt{2} g_a \lambda^{a \dagger} (\psi^\dagger T^a \phi) + \dotsb ,
\end{equation}
where $\lambda^a$ is the SSM gaugino $\tilde g$.  Similar terms also exist for $\tilde \Phi$.  The corresponding vertices are shown in Figure~\ref{fg:1}.  Gauginos obtain masses from one-loop Feymann diagrams as shown in Figure~\ref{fg:2}, which can be calculated by the wave-function renormalization method~\cite{Giudice:1997ni}.  The result is given in \eqref{eq:1-2}.

\begin{fmffile}{fg1}
\end{fmffile}

\begin{fmffile}{fg2}
\end{fmffile}

Generally, messengers can have explicit mass terms as discussed in (extra)ordinary gauge mediation models~\cite{Cheung:2007es, Marques:2009yu}.  The corresponding superpotential is
\begin{equation}
    W = \lambda_{i j} X \tilde \Phi_i \Phi_j + m_{i j} \tilde \Phi_i \Phi_j .
\end{equation}
The resulting gaugino mass is
\begin{equation} \label{eq:2-1}
    M_{\tilde g} \sim \frac{\alpha}{4 \pi} F_X \partial_X \log \det (\lambda X + m) .
\end{equation}
As a consequence of the R-symmetry, it can be shown that the messenger mass matrix has the identity
\begin{equation}
    \det (\lambda X + m) = X^n G(m, \lambda) , \quad
    n = r_X^{-1} \sum_i (2 - R(\tilde \Phi_i) - R(\Phi_i)) .
\end{equation}
So the gaugino mass turns out to be
\begin{equation} \label{eq:2-2}
    M_{\tilde g} \sim \frac{\alpha}{4 \pi} \frac{n F_X}{\langle X \rangle}
\end{equation}
which is similar to the result \eqref{eq:1-2}.

The factor $n$ in \eqref{eq:2-2} is responsible for the vanishing of gaugino masses in a wide class of tree-level SUSY and R-symmetry breaking models, because a classically stable pseudomoduli spaces require $n = 0$~\cite{Komargodski:2009jf}.  One may relax the classically stable condition and build a more general tree-level R-symmetry breaking model.  Although the obstacle can be avoided, \eqref{eq:2-2} only involves $F_X$ and $\langle X \rangle$ which are VEVs of components of the same superfield $X$.  Thus the SUSY breaking pseudomodulus and the R-symmetry breaking field are identical, and \eqref{eq:2-2} is actually a result of loop-level R-symmetry breaking.  To properly study the effect of tree-level R-symmetry breaking, one should refine the model to exclude influence from loop-level R-symmetry breaking effects, as we are to do in the next section.

\section{No-go with separated SUSY and R-symmetry breaking}

\begin{figure}
        \centering 
        \begin{fmfgraph*}(100, 35)
            \fmfleft{i1}
            \fmfright{o1}
            \fmf{fermion,label=$\tilde \psi_i$,label.side=left}{v1,i1}
            \fmf{fermion,label=$\psi_j$,label.side=right}{v1,o1}
            \fmfv{decoration.shape=cross,decoration.size=5thick,label=$i \kappa_{i j} \langle Y \rangle$,label.angle=90}{v1}
        \end{fmfgraph*}
        \qquad
        \begin{fmfgraph*}(100, 35)
            \fmfleft{i1}
            \fmfright{o1}
            \fmf{fermion,label=$\tilde \psi_i^\dagger$,label.side=right}{i1,v1}
            \fmf{fermion,label=$\psi_j^\dagger$,label.side=left}{o1,v1}
            \fmfv{decoration.shape=cross,decoration.size=5thick,label=$i \kappa_{i j}^* \langle Y \rangle^*$,label.angle=90}{v1}
        \end{fmfgraph*}
        \\
        \begin{fmfgraph*}(100, 35)
            \fmfleft{i1}
            \fmfright{o1}
            \fmf{scalar,label=$\phi_i$,label.side=left}{v1,i1}
            \fmf{scalar,label=$\tilde \phi_j$,label.side=right}{v1,o1}
            \fmfv{decoration.shape=cross,decoration.size=5thick,label=$i \lambda_{i j} F_X$,label.angle=90}{v1}
        \end{fmfgraph*}
        \qquad
        \begin{fmfgraph*}(100, 35)
            \fmfleft{i1}
            \fmfright{o1}
            \fmf{scalar,label=$\phi_i^*$,label.side=right}{i1,v1}
            \fmf{scalar,label=$\tilde \phi_j^*$,label.side=left}{o1,v1}
            \fmfv{decoration.shape=cross,decoration.size=5thick,label=$i \lambda_{i j}^* F_X^*$,label.angle=90}{v1}
        \end{fmfgraph*}
        \\[1ex]
        \begin{fmfgraph*}(100, 70)
            \fmfleft{i1}
            \fmfright{o1,o2}
            \fmf{boson,label=$\lambda^a (\tilde g)$,label.side=right}{i1,v1}
            \fmf{fermion}{i1,v1}
            \fmf{scalar,label=$\phi_i^* (\tilde \phi_i^*)$,label.side=right}{v1,o1}
            \fmf{fermion,label=$\psi_i (\tilde \psi_i)$,label.side=right}{o2,v1}
            \fmfv{decoration.shape=circle,decoration.size=2thick,label=$- i \sqrt{2} g_a T^a$,label.angle=0,label.dist=10thin}{v1}
        \end{fmfgraph*}
        \qquad
        \begin{fmfgraph*}(100, 70)
            \fmfleft{i1}
            \fmfright{o1,o2}
            \fmf{boson,label=$\lambda^{a \dagger} (\tilde g^\dagger)$,label.side=left}{v1,i1}
            \fmf{fermion}{v1,i1}
            \fmf{scalar,label=$\phi_i (\tilde \phi_i)$,label.side=left}{o1,v1}
            \fmf{fermion,label=$\psi_i^\dagger (\tilde \psi_i^\dagger)$,label.side=left}{v1,o2}
            \fmfv{decoration.shape=circle,decoration.size=2thick,label=$- i \sqrt{2} g_a T^a$,label.angle=0,label.dist=10thin}{v1}
        \end{fmfgraph*}
        \caption{Messenger coupling vertices related to the gaugino mass.}
        \label{fg:3}
    \end{figure}
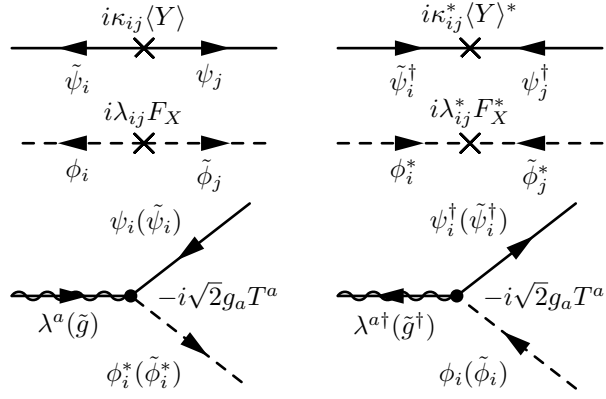

The essential concept of tree-level R-symmetry breaking is the misalignment between the SUSY breaking pseudomodulus and the R-symmetry breaking field~\cite{Carpenter:2008wi, Sun:2008va, Komargodski:2009jf}.  The R-symmetry breaking field can be decomposed to a component parallel to the pseudomodulus which actually comes from loop-level R-symmetry breaking, and a transverse component which really counts for tree-level R-symmetry breaking.  Based on this decomposition, we suppose that there are two separated fields $X$ and $Y$ which respectively break SUSY and the R-symmetry.  At the vacuum we have
\begin{equation} \label{eq:3-1}
    \langle X \rangle = 0 , \quad
    F_X \ne 0 , \quad
    \langle Y \rangle \ne 0 , \quad
    F_Y = 0 ,
\end{equation}
which is just \eqref{eq:1-3} in the form of component fields.  The rest of our derivation follows the standard SUSY formulation as shown before.  Both $X$ and $Y$ couple to messengers through cubic terms in the superpotential and give the messenger spectrum
\begin{align}
    [\lambda_{i j} X \tilde \Phi_i \Phi_j]_{\theta \theta} + \text{c.c.} &= \lambda_{i j} F_X \tilde \phi_i \phi_j + \lambda_{i j}^* F_X^* \tilde \phi_i^* \phi_j^* + \dotsb ,\\
    [\kappa_{i j} Y \tilde \Phi_i \Phi_j]_{\theta \theta} + \text{c.c.} &= \kappa_{i j} \langle Y \rangle \tilde \psi_i \psi_j + \kappa_{i j}^* \langle Y \rangle^* \tilde \psi_i^\dagger \psi_j^\dagger + \dotsb .
\end{align}
Clashing vertices similar to the ones addressed before can be determined from these expressions.  The contact terms of messengers and gauginos are 
\begin{equation}
    [\Phi_i^\dagger (e^{2 g_a T^a V}) \Phi_i]_{\theta \theta \bar \theta \bar \theta} = - \sqrt{2} g_a (\phi_i^* T^a \psi_i) \lambda^a - \sqrt{2} g_a \lambda^{a \dagger} (\psi_i^\dagger T^a \phi_i) + \text{c.c.} + \dotsb ,
\end{equation}
where a minimal K\"ahler potential is taken to keep the R-symmetry.  The corresponding vertices are shown in Figure~\ref{fg:3}.

\begin{fmffile}{fg3}
\end{fmffile}

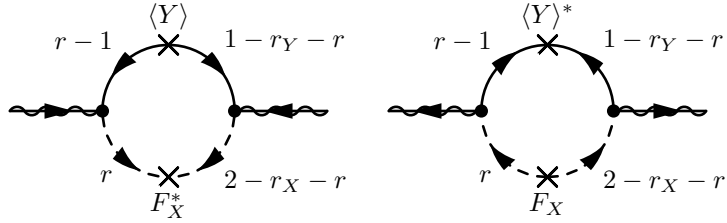
\begin{figure}
        \centering
        \begin{fmfgraph*}(120,80)
            \fmfleft{i1}
            \fmfright{o1}
            \fmf{fermion}{i1,v1}
            \fmf{photon}{i1,v1}
            \fmf{fermion}{o1,v2}
            \fmf{photon}{o1,v2}
            \fmf{phantom,left,tension=0.7,tag=1}{v1,v2}
            \fmf{phantom,right,tension=0.7,tag=2}{v1,v2}
            \fmfdot{v1,v2}
            \fmffreeze
            \fmfipath{p[]}
            \fmfiset{p1}{vpath1(__v1,__v2)}
            \fmfiset{p2}{vpath2(__v1,__v2)}
            \fmfi{fermion,label=$r - 1$}{subpath (length(p1)/2,0) of p1}
            \fmfi{fermion,label=$1 - r_Y - r$}{subpath (length(p1)/2,length(p1)) of p1}
            \fmfi{scalar,label=$r$}{subpath (0,length(p2)/2) of p2}
            \fmfi{scalar,label=$2 - r_X - r$}{subpath (length(p2),length(p2)/2) of p2}
            \fmfiv{decoration.shape=cross,decoration.size=5thick,label=$\langle Y \rangle$,label.angle=90}{point length(p1)/2 of p1}
            \fmfiv{decoration.shape=cross,decoration.size=5thick,label=$F_X^*$,label.angle=-90}{point length(p2)/2 of p2}
        \end{fmfgraph*}
        \qquad
        \begin{fmfgraph*}(120,80)
            \fmfleft{i1}
            \fmfright{o1}
            \fmf{fermion}{v1,i1}
            \fmf{photon}{v1,i1}
            \fmf{fermion}{v2,o1}
            \fmf{photon}{v2,o1}
            \fmf{phantom,left,tension=0.7,tag=1}{v1,v2}
            \fmf{phantom,right,tension=0.7,tag=2}{v1,v2}
            \fmfdot{v1,v2}
            \fmffreeze
            \fmfipath{p[]}
            \fmfiset{p1}{vpath1(__v1,__v2)}
            \fmfiset{p2}{vpath2(__v1,__v2)}
            \fmfi{fermion,label=$r - 1$}{subpath (0,length(p1)/2) of p1}
            \fmfi{fermion,label=$1 - r_Y - r$}{subpath (length(p1),length(p1)/2) of p1}
            \fmfi{scalar,label=$r$}{subpath (length(p2)/2,0) of p2}
            \fmfi{scalar,label=$2 - r_X - r$}{subpath (length(p2)/2,length(p2)) of p2}
            \fmfiv{decoration.shape=cross,decoration.size=5thick,label=$\langle Y \rangle^*$,label.angle=90}{point length(p1)/2 of p1}
            \fmfiv{decoration.shape=cross,decoration.size=5thick,label=$F_X$,label.angle=-90}{point length(p2)/2 of p2}
        \end{fmfgraph*}
        \caption{Possible one-loop diagrams for the gaugino mass from tree-level R-symmetry breaking, with R-charges of messenger components determined from R-charge conservation.  The two diagrams are conjugate to each other, and both should be included as Majorana mass terms.}
        \label{fg:4}
    \end{figure}

At first glance, one can draw diagrams as shown in Figure~\ref{fg:4} which give a similar gaugino mass as done in ordinary gauge mediation models.  But after checking R-charge conservation conditions for each vertex in the loop, it is found that the loop diagrams in Figure~\ref{fg:4} are valid only if $r_X = r_Y$.  Then the R-symmetry allows $X$ to have all the messenger couplings which $Y$ has, and vise versa.  There is no clear distinction between $X$ and $Y$ fields and the separation \eqref{eq:3-1} becomes non-generic.

\begin{fmffile}{fg4}
\end{fmffile}

In generic models with $r_X = r_Y$, all the four VEVs in \eqref{eq:3-1} are non-zero.  If we make a linear combination of them aligned with the F-term, the combination field generically has non-zero VEV.  Thus tree-level R-symmetry breaking, if existing in this type of models, is accompanied with the same magnitude of loop-level R-symmetry breaking contributing to the gaugino mass which makes tree-level R-symmetry breaking redundant.  For models with $r_X \ne r_Y$, $X$ and $Y$ fields are clearly separated, \eqref{eq:3-1} can be naturally satisfied and tree-level R-symmetry is well defined without interference from loop-level R-symmetry breaking.  But as we have shown, there is no valid one-loop diagram for the gaugino mass in this case.

\section{Bypassing the no-go with explicit messenger mass terms}

\begin{figure}
        \centering
        \begin{fmfgraph*}(210,160)
            \fmfleft{i1}
            \fmfright{o1}
            \fmf{fermion}{i1,v1}
            \fmf{photon}{i1,v1}
            \fmf{fermion}{o1,v2}
            \fmf{photon}{o1,v2}
            \fmf{phantom,left,tension=0.3,tag=1}{v1,v2}
            \fmf{phantom,right,tension=0.3,tag=2}{v1,v2}
            \fmfdot{v1,v2}
            \fmffreeze
            \fmfipath{p[]}
            \fmfiset{p1}{vpath1(__v1,__v2)}
            \fmfiset{p2}{vpath2(__v1,__v2)}
            \fmfi{fermion,label=$r - 1$}{subpath (length(p1)/6,0) of p1}
            \fmfi{fermion,label=$1 - r_Y - r$}{subpath (length(p1)/6,length(p1)/3) of p1}
            \fmfi{fermion,label=$r - 1$}{subpath (length(p1)/2,length(p1)/3) of p1}
            \fmfi{dots}{subpath (length(p1)/2,5*length(p1)/6) of p1}
            \fmfi{fermion,label=$1 - r_Y - r$}{subpath (5*length(p1)/6,length(p1)) of p1}
            \fmfi{scalar,label=$r$}{subpath (0,length(p2)/6) of p2}
            \fmfi{scalar,label=$2 - r_X - r$}{subpath (length(p2)/3,length(p2)/6) of p2}
            \fmfi{scalar,label=$r$}{subpath (length(p2)/3,length(p2)/2) of p2}
            \fmfi{dots}{subpath (length(p2)/2,5*length(p2)/6) of p2}
            \fmfi{scalar,label=$2 - r_X - r$}{subpath (length(p2),5*length(p2)/6) of p2}
            \fmfiv{decoration.shape=cross,decoration.size=5thick,decoration.angle=60,label=$\langle Y \rangle$}{point length(p1)/6 of p1}
            \fmfiv{decoration.shape=cross,decoration.size=5thick,decoration.angle=30,label=$\langle Y \rangle^*$}{point length(p1)/3 of p1}
            \fmfiv{decoration.shape=cross,decoration.size=5thick,label=$\langle Y \rangle$}{point length(p1)/2 of p1}
            \fmfiv{decoration.shape=cross,decoration.size=5thick,decoration.angle=-60,label=$F_X^*$}{point length(p2)/6 of p2}
            \fmfiv{decoration.shape=cross,decoration.size=5thick,decoration.angle=-30,label=$F_X$}{point length(p2)/3 of p2}
            \fmfiv{decoration.shape=cross,decoration.size=5thick,label=$F_X^*$}{point length(p2)/2 of p2}
        \end{fmfgraph*}
        \qquad
        \begin{fmfgraph*}(210,160)
            \fmfleft{i1}
            \fmfright{o1}
            \fmf{fermion}{v1,i1}
            \fmf{photon}{v1,i1}
            \fmf{fermion}{v2,o1}
            \fmf{photon}{v2,o1}
            \fmf{phantom,left,tension=0.3,tag=1}{v1,v2}
            \fmf{phantom,right,tension=0.3,tag=2}{v1,v2}
            \fmfdot{v1,v2}
            \fmffreeze
            \fmfipath{p[]}
            \fmfiset{p1}{vpath1(__v1,__v2)}
            \fmfiset{p2}{vpath2(__v1,__v2)}
            \fmfi{fermion,label=$r - 1$}{subpath (0,length(p1)/6) of p1}
            \fmfi{fermion,label=$1 - r_Y - r$}{subpath (length(p1)/3,length(p1)/6) of p1}
            \fmfi{fermion,label=$r - 1$}{subpath (length(p1)/3,length(p1)/2) of p1}
            \fmfi{dots}{subpath (length(p1)/2,5*length(p1)/6) of p1}
            \fmfi{fermion,label=$1 - r_Y - r$}{subpath (length(p1),5*length(p1)/6) of p1}
            \fmfi{scalar,label=$r$}{subpath (length(p2)/6,0) of p2}
            \fmfi{scalar,label=$2 - r_X - r$}{subpath (length(p2)/6,length(p2)/3) of p2}
            \fmfi{scalar,label=$r$}{subpath (length(p2)/2,length(p2)/3) of p2}
            \fmfi{dots}{subpath (length(p2)/2,5*length(p2)/6) of p2}
            \fmfi{scalar,label=$2 - r_X - r$}{subpath (5*length(p2)/6,length(p2)) of p2}
            \fmfiv{decoration.shape=cross,decoration.size=5thick,decoration.angle=60,label=$\langle Y \rangle^*$}{point length(p1)/6 of p1}
            \fmfiv{decoration.shape=cross,decoration.size=5thick,decoration.angle=30,label=$\langle Y \rangle$}{point length(p1)/3 of p1}
            \fmfiv{decoration.shape=cross,decoration.size=5thick,label=$\langle Y \rangle^*$}{point length(p1)/2 of p1}
            \fmfiv{decoration.shape=cross,decoration.size=5thick,decoration.angle=-60,label=$F_X$}{point length(p2)/6 of p2}
            \fmfiv{decoration.shape=cross,decoration.size=5thick,decoration.angle=-30,label=$F_X^*$}{point length(p2)/3 of p2}
            \fmfiv{decoration.shape=cross,decoration.size=5thick,label=$F_X$}{point length(p2)/2 of p2}
        \end{fmfgraph*}
        \caption{Possible one-loop diagrams for the gaugino mass with multiple vertices inserted, with R-charges of messenger components determined from R-charge conservation.  The two diagrams are conjugate to each other, and both should be included as Majorana mass terms.}
        \label{fg:5}
    \end{figure}
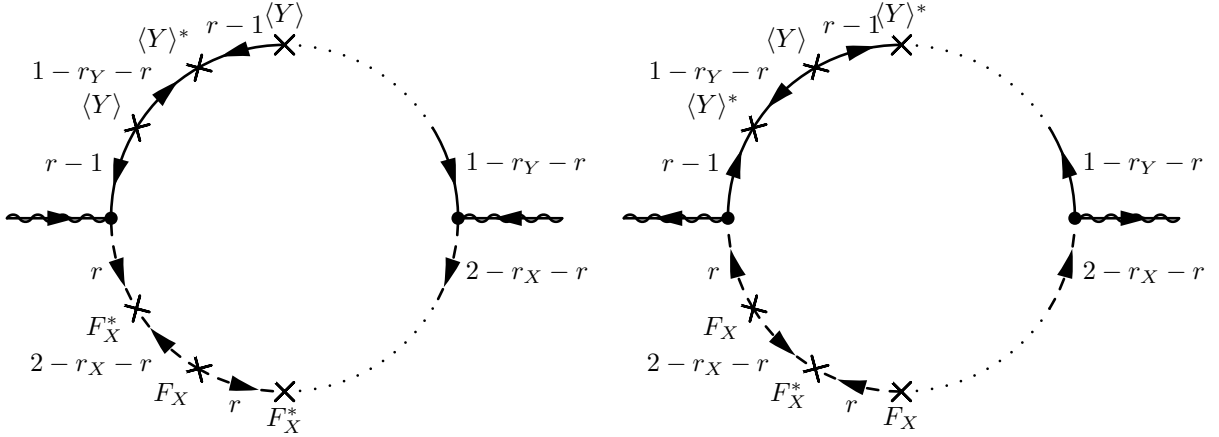

To see whether the gaugino mass can be generated beyond our simplest type of models, one may try to insert more spurion vertices into the loop, as shown in Figure~\ref{fg:5}.  Noticing directions of propagators, the Feynman rules require that both the fermion line and the boson line have odd numbers of vertices inserted.  Checking the R-symmetry conservation condition turns out that $r_X = r_Y$ is still required for a valid diagram.  So we get the same no-go conclusion as discussed before.

\begin{fmffile}{fg5}
\end{fmffile}

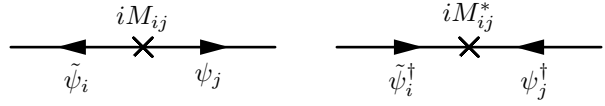
\begin{figure}
        \centering 
        \begin{fmfgraph*}(100, 35)
            \fmfleft{i1}
            \fmfright{o1}
            \fmf{fermion,label=$\tilde \psi_i$,label.side=left}{v1,i1}
            \fmf{fermion,label=$\psi_j$,label.side=right}{v1,o1}
            \fmfv{decoration.shape=cross,decoration.size=5thick,label=$i M_{i j}$,label.angle=90}{v1}
        \end{fmfgraph*}
        \qquad
        \begin{fmfgraph*}(100, 35)
            \fmfleft{i1}
            \fmfright{o1}
            \fmf{fermion,label=$\tilde \psi_i^\dagger$,label.side=right}{i1,v1}
            \fmf{fermion,label=$\psi_j^\dagger$,label.side=left}{o1,v1}
            \fmfv{decoration.shape=cross,decoration.size=5thick,label=$i M_{i j}^*$,label.angle=90}{v1}
        \end{fmfgraph*}
        \caption{Messenger clashing vertices from explicit mass terms.}
        \label{fg:6}
    \end{figure}

One feature of the diagram in Figure~\ref{fg:5} is the alternation of messenger component R-charges between $r - 1$ and $1 - r_Y - r$ on the fermion line, or between $r$ and $2 - r_Y - r$ on the boson line.  So one way one may bypass the no-go theorem is to consider loop diagrams with a different R-charge pattern.  This is possible by introducing explicit messenger mass terms in the superpotential.  Expanding mass terms in components gives
\begin{equation}
    [M_{i j} \tilde \Phi_i \Phi_j]_{\theta \theta} + \text{c.c.} = M_{i j} \tilde \psi_i \psi_j + M_{i j}^* \tilde \psi_i^\dagger \psi_j^\dagger
\end{equation}
and new clashing vertices as shown in Figure~\ref{fg:6}.  Since mass parameters $M_{i j}$ do not carry R-charges, each two fermion components joining such a vertex should have opposite R-charges.  Inserting these mass vertices as well as $Y$ vertices into the fermion line, and recalling that $Y$ has a non-zero R-charge to break the R-symmetry, a loop diagram similar to Figure~\ref{fg:5} may be valid with a particular R-charge arrangement, and the previous no-go statement may be bypassed.

\begin{fmffile}{fg6}
\end{fmffile}

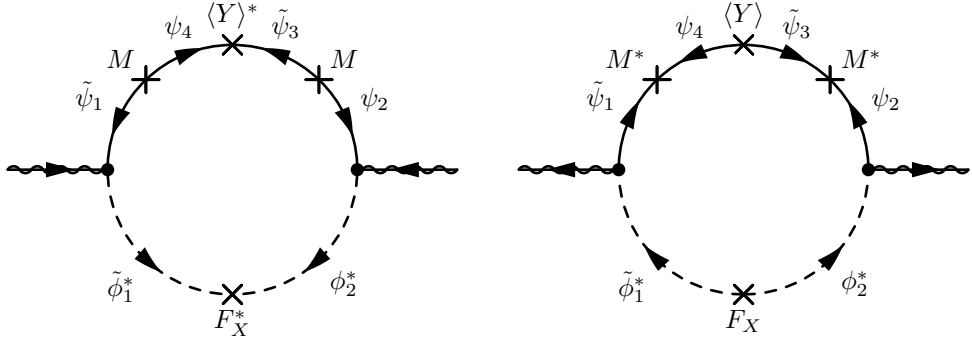
\begin{figure}
        \centering
        \begin{fmfgraph*}(170,120)
            \fmfleft{i1}
            \fmfright{o1}
            \fmf{fermion}{i1,v1}
            \fmf{photon}{i1,v1}
            \fmf{fermion}{o1,v2}
            \fmf{photon}{o1,v2}
            \fmf{phantom,left,tension=0.4,tag=1}{v1,v2}
            \fmf{phantom,right,tension=0.4,tag=2}{v1,v2}
            \fmfdot{v1,v2}
            \fmffreeze
            \fmfipath{p[]}
            \fmfiset{p1}{vpath1(__v1,__v2)}
            \fmfiset{p2}{vpath2(__v1,__v2)}
            \fmfi{fermion,label=$\tilde \psi_1$}{subpath (length(p1)/4,0) of p1}
            \fmfi{fermion,label=$\psi_4$}{subpath (length(p1)/4,length(p1)/2) of p1}
            \fmfi{fermion,label=$\tilde \psi_3$}{subpath (3*length(p1)/4,length(p1)/2) of p1}
            \fmfi{fermion,label=$\psi_2$}{subpath (3*length(p1)/4,length(p1)) of p1}
            \fmfi{scalar,label=$\tilde \phi_1^*$}{subpath (0,length(p2)/2) of p2}
            \fmfi{scalar,label=$\phi_2^*$}{subpath (length(p2),length(p2)/2) of p2}
            \fmfiv{decoration.shape=cross,decoration.size=5thick,decoration.angle=45,label=$M$}{point length(p1)/4 of p1}
            \fmfiv{decoration.shape=cross,decoration.size=5thick,label=$\langle Y \rangle^*$}{point length(p1)/2 of p1}
            \fmfiv{decoration.shape=cross,decoration.size=5thick,decoration.angle=-45,label=$M$}{point 3*length(p1)/4 of p1}
            \fmfiv{decoration.shape=cross,decoration.size=5thick,label=$F_X^*$}{point length(p2)/2 of p2}
        \end{fmfgraph*}
        \qquad
        \begin{fmfgraph*}(170,120)
            \fmfleft{i1}
            \fmfright{o1}
            \fmf{fermion}{v1,i1}
            \fmf{photon}{v1,i1}
            \fmf{fermion}{v2,o1}
            \fmf{photon}{v2,o1}
            \fmf{phantom,left,tension=0.4,tag=1}{v1,v2}
            \fmf{phantom,right,tension=0.4,tag=2}{v1,v2}
            \fmfdot{v1,v2}
            \fmffreeze
            \fmfipath{p[]}
            \fmfiset{p1}{vpath1(__v1,__v2)}
            \fmfiset{p2}{vpath2(__v1,__v2)}
            \fmfi{fermion,label=$\tilde \psi_1$}{subpath (0,length(p1)/4) of p1}
            \fmfi{fermion,label=$\psi_4$}{subpath (length(p1)/2,length(p1)/4) of p1}
            \fmfi{fermion,label=$\tilde \psi_3$}{subpath (length(p1)/2,3*length(p1)/4) of p1}
            \fmfi{fermion,label=$\psi_2$}{subpath (length(p1),3*length(p1)/4) of p1}
            \fmfi{scalar,label=$\tilde \phi_1^*$}{subpath (length(p2)/2,0) of p2}
            \fmfi{scalar,label=$\phi_2^*$}{subpath (length(p2)/2,length(p2)) of p2}
            \fmfiv{decoration.shape=cross,decoration.size=5thick,decoration.angle=45,label=$M^*$}{point length(p1)/4 of p1}
            \fmfiv{decoration.shape=cross,decoration.size=5thick,label=$\langle Y \rangle$}{point length(p1)/2 of p1}
            \fmfiv{decoration.shape=cross,decoration.size=5thick,decoration.angle=-45,label=$M^*$}{point 3*length(p1)/4 of p1}
            \fmfiv{decoration.shape=cross,decoration.size=5thick,label=$F_X$}{point length(p2)/2 of p2}
        \end{fmfgraph*}
        \caption{One-loop diagrams for the gaugino mass from the example model \eqref{eq:4-1}, with R-charge conservation at every vertices.  Diagrams on the left and right are conjugate to each other, and both should be included as Majorana mass terms.}
        \label{fg:7}
    \end{figure}

The possibility of bypassing the no-go theorem can be demonstrated in the following example with the superpotential
\begin{equation} \label{eq:4-1}
    W = \lambda X \tilde \Phi_1 \Phi_2 + \kappa Y \tilde \Phi_3 \Phi_4 + M_1 \tilde \Phi_1 \Phi_4 + M_2 \tilde \Phi_3 \Phi_2 .
\end{equation}
Following our previous convention of notation, $X$ and $Y$ are SUSY and R-symmetry breaking spurions, and messengers with and without tildes are conjugate to each other in SM gauge symmetries.  R-charges of messenger superfields can be consistently assigned as
\begin{equation}
    r_X = - r_Y , \quad
    r_2 = 2 - r_X - r_1 , \quad
    r_3 = r_X + r_1 , \quad
    r_4 = 2 - r_1 ,
\end{equation}
where the free parameters $r_X$ and $r_1$ may be fixed by some UV dynamics.  For simplicity we assume $\lambda \sim \kappa \sim 1$ and $M \sim M_1 \sim M_2 \gtrsim \langle Y \rangle$, so all fermion and boson components of messengers have mass around $M$.  A loop diagram for the gaugino mass can be obtained from previous Feynman rules with R-charge conservation at all vertices, as shown in Figure~\ref{fg:7}.  Up to an order $1$ overall factor, the loop diagram in Figure~\ref{fg:7} evaluates as
\begin{equation} \label{eq:4-2}
    M_{\tilde g} \sim \int \frac{\ud p^4}{(2 \pi)^4} \frac{2 g_a^2 \kappa \lambda M_1 M_2 F_X \langle Y \rangle}{(p^2 - M^2)^2 (\gamma^\mu p_\mu - M)^4}
                 \sim \frac{\alpha}{4 \pi} \frac{F_X \langle Y \rangle}{M^2} .
\end{equation}

\begin{fmffile}{fg7}
\end{fmffile}

Alternatively, the gaugino mass can be obtained from the wave-function renormalization method calculation result \eqref{eq:2-1}.  The model \eqref{eq:4-1} has the messenger mass matrix with determinant
\begin{equation}
    \det \begin{pmatrix}
             0         & \lambda X & 0        & M_1\\
             \lambda X & 0         & M_2      & 0\\
             0         & M_2       & 0        & \kappa Y\\
             M_1       & 0         & \kappa Y & 0\\ 
         \end{pmatrix}
    = (M_1 M_2 - \kappa \lambda X Y)^2
    \sim (M^2 - X Y)^2 .
\end{equation}
Noticing $\langle X \rangle = 0$, the gaugino mass turns out to be
\begin{equation} \label{eq:4-3}
    M_{\tilde g} \sim \frac{\alpha}{4 \pi} F_X \partial_X \log (M^2 - X Y)^2
                 \sim \frac{\alpha}{4 \pi} \frac{F_X \langle Y \rangle}{M^2} .
\end{equation}

Although both \eqref{eq:4-2} and \eqref{eq:4-3} give consistent results for the non-varnishing one-loop gaugino mass, there is a suppression from the messenger mass scale $M$, which is supposed to be larger than the scales of both SUSY breaking $F_X$ and R-symmetry breaking $\langle Y \rangle$.  Generically, if it is possible to generate the gaugino mass by this means, a number of mass vertices need to be inserted into the fermion line, and the same number of fermion propagators need to be added in the loop accordingly.  Then the calculation \eqref{eq:4-2} gives a similar answer to \eqref{eq:4-3}, that the gaugino mass is always suppressed by the ratio $\langle Y \rangle / M$ comparing to the loop-level R-symmetry breaking result \eqref{eq:1-2}.

\section{Conclusion and outlook}

We have shown that with a simplified assumption of tree-level R-symmetry breaking where SUSY and R-symmetries are broken by different fields, the gaugino mass either becomes zero at one loop or gets contribution from loop-level R-symmetry breaking.  So tree-level R-symmetry breaking either fails its original motivation to generate the gaugino mass, or becomes unnecessary because of the existence of loop-level R-symmetry breaking.  Thus tree-level R-symmetry breaking is proved to be either no-go or redundant for phenomenology in such simple models.  Including messenger mass terms in the superpotential and inserting mass vertices into the fermion line may be helpful to bypass the no-go theorem.  Our simple argument shows that with a particular R-charge arrangement, the resulting gaugino mass has the generic form
\begin{equation}
    M_{\tilde g} \sim \frac{\alpha}{4 \pi} \frac{F_X \langle Y \rangle}{M^2} ,
\end{equation}
indicating a suppression by the ratio between the R-symmetry breaking scale and the messenger mass scale.

The essential assumption in our proof is the separation of SUSY and R-symmetry breaking fields as in \eqref{eq:3-1}.  So our analysis covers a wider range of models than just the tree-level R-symmetry breaking case, such as in the Goldstini scenario~\cite{Cheung:2010mc}.  Whenever there are separated SUSY breaking sector and R-symmetry breaking sector, all the proof can be followed and we have a similar no-go statement.  For simplicity, it is well enough to obtain the gaugino mass from loop-level R-symmetry breaking from a single spurion, unless other phenomenology features require a multi-spurion model.

It should be addressed that our work deals with the Majorana gaugino mass which is one of the motivations of R-symmetry breaking.  There are alternative models proposing Dirac gaugino mass terms, which can be generated from either D-terms or R-symmetric F-terms~\cite{Fayet:1978qc, Hall:1990hq, Fox:2002bu, Benakli:2008pg}.  Besides the well-studied Majorana gaugino mass from loop-level R-symmetry breaking, there are various models and features outside of our no-go statement for one to explore.

\section*{Acknowledgement}

We thank Pierre Fayet, Tianjun Li and Dimitri Polyakov for helpful discussions.  This work is supported by the National Natural Science Foundation of China under grant 11305110.

\end{document}